\documentclass[prl,aps,a4paper,10pt,twocolumn,superscriptaddress,amsmath,amssymb,showpacs]{revtex4}
\usepackage{bm}
\usepackage{epsfig}
\usepackage{amsmath}

\renewcommand{\v}[1]{{\bf #1}}
\newcommand{\bpm}{\begin{pmatrix}}
\newcommand{\epm}{\end{pmatrix}}
\newcommand{\ba}{\begin{eqnarray}}
\newcommand{\ea}{\end{eqnarray}}
\newcommand{\nn}{\nonumber \\}

\begin{document}

\title{Emergence of Orbital Angular Momentum by Inversion Symmetry Breaking \\
and Its Detection by ARPES}
\author{Choong H. Kim} \affiliation{Department of
Physics and Astronomy, Seoul National University, Seoul 151-742,
Korea}

\author{Jin-Hong Park} \affiliation{Department of Physics and
BK21 Physics Research Division, Sungkyunkwan University, Suwon
440-746, Korea}
\author{Jun Won Rhim}
\affiliation{School of Physics, Korea Institute for Advanced Study,
Seoul 130-722, Korea}
\author{Beom Young Kim}\affiliation{Institute of Physics and Applied Physics,
Yonsei University, Seoul, Korea}
\author{Jaejun Yu}\affiliation{Department of Physics and
Astronomy, Seoul National University, Seoul 151-742, Korea}
\author{Masashi Arita}\affiliation{Hiroshima Synchrotron Radiation Center,
Hiroshima University, Higashi-Hiroshima, Hiroshima 739-0046, Japan}
\author{Kenya Shimada}\affiliation{Hiroshima Synchrotron Radiation Center,
Hiroshima University, Higashi-Hiroshima, Hiroshima 739-0046, Japan}
\author{Hirofumi Namatame}\affiliation{Hiroshima Synchrotron Radiation Center,
Hiroshima University, Higashi-Hiroshima, Hiroshima 739-0046, Japan}
\author{Masaki Taniguchi}\affiliation{Hiroshima Synchrotron Radiation Center,
Hiroshima University, Higashi-Hiroshima, Hiroshima 739-0046, Japan}
\author{Changyoung Kim}\email{changyoung@yonsei.ac.kr}\affiliation{Institute of Physics and Applied Physics,
Yonsei University, Seoul, Korea}
\author{Jung Hoon Han}\email{hanjh@skku.edu}\affiliation{Department of Physics and
BK21 Physics Research Division, Sungkyunkwan University, Suwon
440-746, Korea}
\date{\today}

\begin{abstract} Rashba-split surface band is characterized by a
one-to-one correspondence between the electron's momentum $\v k$ and
its spin orientation. Here we show that a similar correspondence
between momentum and orbital angular momentum (OAM) must exist on
surface bands once the inversion symmetry is broken. The
correspondence is valid even when there is no spin-orbit interaction.
Tight-binding and first-principles calculations are presented to
support our claim. As a method to detect such OAM-momentum
correspondence, we propose the circular dichroism (CD) experiment
using the angle-resolved photoemission (ARPES) setup. CD-ARPES
experiment performed on Cu surface confirms the existence of chiral
OAM. A new concept of ``orbital Galvanic effect" is
proposed.\end{abstract}

\maketitle

Electrons in solids are classified by several quantum numbers,
including their momentum, spin, and sometimes orbital angular
momentum (OAM). In some cases two or more degrees of freedom appear
in coupled form as, for instance, in the Rashba phenomena where
there exists a one-to-one correspondence between the electron's
momentum and its spin\cite{rashba}. Rashba effects are prominent in
certain surface bands\cite{bihlmayer-review} or in a bulk
insulator\cite{BTI} lacking the inversion symmetry. In addition to
inversion symmetry breaking (ISB), spin-orbit interaction (SOI) is
the pre-requisite for the observability of Rashba-related phenomena.
The relation of SOI to Rashba splitting has been discussed by
several authors in the past\cite{petersen}, and recently by some of
the present authors\cite{park-Rashba}.

The implication of ISB on the Rashba splitting is anticipated on
symmetry grounds. With the surface normal along the
$\hat{z}$-direction, one can write down a symmetry-allowed
Hamiltonian $H_\mathrm{R} = \lambda_\mathrm{R} \hat{z} \cdot (\v k
\times \bm \sigma )$ in terms of the spin operator $\bm \sigma/2$
and electronic momentum $\v k$. The Rashba energy scale
$\lambda_\mathrm{R}$ appearing in $H_\mathrm{R}$ is on the order of
the electrostatic potential barrier across the surface, multiplied
by its width\cite{park-Rashba}. The chiral spin angular momentum
(SAM) structure in momentum space follows as a direct consequence of
the Rashba Hamiltonian $H_\mathrm{R}$. Such chiral SAM structure has
been thoroughly documented on surfaces of several metallic
elements\cite{SARPES,kimura} as well as on the surfaces of
topological insulators\cite{kane-review} in recent years. Its
potential as an effective source of spin current is also being
actively investigated\cite{kimura}.

Upon closer inspection, however, one finds that other physical
quantities with the same symmetry properties can take the place of
$\bm \sigma$ in the Rashba Hamiltonian. It is conceivable, for
instance, to replace $\bm \sigma$ by the orbital angular momentum
operator $\bm L$ in a degenerate orbital system where $\bm L$ is an
active degree of freedom. In the case of materials with strong
spin-orbit interaction (SOI), the total angular momentum $\bm J =
\bm L + (1/2)\bm \sigma$ may be the more appropriate
choice\cite{park-Rashba}. Both scenarios would imply chiral OAM
structure in much the same way that chiral SAM follows from the
Rashba Hamiltonian. We show that chiral OAM in one-to-one
correspondence with the electron's linear momentum is indeed a
general consequence of ISB at the surface and should occur on
surface bands even for materials with no SOI, placing OAM as the
more generic feature of ISB than SAM is.

The above symmetry consideration is further supported by the
analysis of a tight-binding (TB) model of two-dimensional monolayer
of atoms. For simplicity we will consider an s$p$-orbital system
with three degenerate $p$-orbitals forming bands. To emphasize the
notion of ISB-induced OAM better, we will first study the spinless
case to show that OAM can arise without SOI and only later introduce
SOI as a perturbation. The latter procedure is shown to recover the
usual Rashba spin splitting. It will be shown that the phenomenon of
chiral OAM is described by a two-dimensional massive Dirac
Hamiltonian.

The tight-binding model of spinless $p$-orbitals can be constructed
in terms of two Slater-Koster parameters $V_1$ and $V_2$ for
$\sigma$- and $\pi$-bonding amplitudes, respectively, and a third
one, $\gamma$, representing the degree of ISB\cite{petersen}. In
real materials, $\gamma$ arises from the surface-normal electric
field which breaks the inversion symmetry. Degeneracy of atomic
$p$-orbital states is assumed in our model, which is justified in
simple elements like Cu, Sb and Bi due to their weak crystal field
splitting. We will write $N$ for the number of sites in the lattice,
and $|i, p_\lambda \rangle$ for the localized Wannier orbitals
$(\lambda=x,y,z$) at the atomic site $\v r_i$. Then the
tight-binding Hamiltonian for the triangular lattice (essentially
the same result obtains for square lattice) near the $\Gamma$-point
($\v k = 0$) in the momentum-space basis $|\v k , p_\lambda \rangle
= N^{-1/2} \sum_i e^{i \v k \cdot \v r_i} | i, p_\lambda \rangle$
becomes

\ba H_{\v k}=\bpm
\alpha k_x^2+\beta k_y^2&(\alpha-\beta)k_xk_y&-i\frac{3}{2}\gamma k_x \\
(\alpha-\beta)k_xk_y &\alpha k_y^2+\beta k_x^2 & -i\frac{3}{2}\gamma
k_y\\i\frac{3}{2}\gamma k_x&i\frac{3}{2}\gamma
k_y&4(\alpha-\beta)-\frac{3}{2}V_2k^2\epm . \label{eq:OAM-model}\ea
Here, $\alpha=3(3V_1-V_2)/8$, $\beta=3(V_1-3V_2)/8$ and
$k^2=k_x^2+k_y^2$. The lattice constant is taken to be unity. To
diagonalize $H_{\v k}$, it is convenient to choose a new set of
basis vectors

\ba |\mathrm{I}, \v k\rangle &=& (k_y /k) |p_x, \v k \rangle -(k_x
/k) |p_y, \v k \rangle, \nn
|\mathrm{II}, \v k\rangle &=& (k_x /k) |p_x, \v k \rangle +(k_y /k)
|p_y, \v k \rangle, \nn
|\mathrm{III}, \v k\rangle &=& e^{-i\phi_{\v k}} |p_z, \v k \rangle,
\label{eq:8} \ea
$k =|\v k |$, $e^{i\phi_{\v k}} = (k_x + ik_y)/k$. The state
$|\mathrm{I}, \v k\rangle$ remains decoupled at energy $E_{1,\v k} =
3 V_2 - 3 V_1 +3(V_1-3 V_2) \v k^2 /8$, while $|\mathrm{II}, \v
k\rangle$ and $|\mathrm{III}, \v k\rangle$ obey a reduced $2\times2$
Hamiltonian

\ba H_\mathrm{OAM} = -4\beta I_{2\times 2} + M^{-1}\v
k^2+\frac{3}{2}\hat{z}\cdot(\gamma\v k\times
\bm{\tau}-\Delta\tau^z), \label{eq:reduced-H}\ea
with $\Delta = V_1 + V_2$ the bandwidth,  $M^{-1}=\bpm \alpha & 0 \\
0 & -3V_2/2\epm$ the effective mass tensor, and $\bm{\tau}$ the
pseudo-spin matrix.  Eigenstates in the leading order of $\gamma
/\Delta$ are

\ba |2, \v k \rangle &\simeq&  |\mathrm{II}, \v k\rangle - {i \gamma
(k_x - ik_y) \over 2\Delta}|\mathrm{III}, \v k \rangle, \nn
|3, \v k \rangle &\simeq &|\mathrm{III}, \v k \rangle - {i\gamma (k_x
+ ik_y) \over 2\Delta}|\mathrm{II}, \v k\rangle , \label{eq:10}\ea
with energies $E_{2, \v k} \simeq 3(V_2-V_1) +3 ((3V_1 - V_2 )/8
-\gamma^2/4\Delta) \v k^2$ and $E_{3, \v k} \simeq 6V_2 +
(3\gamma^2/4\Delta-3V_2/2) \v k^2$. The OAM operator is given by the
sum $\bm L = (1/N) \sum_i \bm L_i$ where each $\bm L_i$ acts on the
Wannier state $|i , p_\lambda\rangle$ as the usual $L=1$ angular
momentum operator. The two bands obtained above carry nonzero OAM as
claimed ($L^+ = L^x + i L^y$):

\ba \langle 2, \v k |L^+| 2, \v k \rangle \!=\! {i \gamma \over
\Delta } (k_x \!+\! i k_y ) \!=\! - \langle 3, \v k |L^+|3, \v k
\rangle . \label{eq:perturbative-OAM}\ea
Both bands possess a chiral pattern of OAM whose strength is
proportional to the ISB parameter $\gamma$ and are opposite between
the two bands. OAM for the decoupled state $|\mathrm{I}, \v
k\rangle$ is vanishing because it consists of $p_x$ and $p_y$
orbitals only ($L_z = \pm 1$) and does not contain $p_z$ ($L_z =
0$). The reduced Hamiltonian (\ref{eq:reduced-H}) satisfies
time-reversal invariance. The size of OAM gets reduced linearly with
the momentum $|\v k|$ and with the symmetry-breaking parameter
$\gamma$. By contrast, the magnitude of polarized spin in the Rashba
model is independent of the Rashba parameter $\lambda_\mathrm{R}$.

Incorporating the spin degree of freedom at each atomic site as the
Pauli matrix $\bm \sigma_i$, we can examine the influence of the SOI
interaction $H_\mathrm{so}=(\lambda_\mathrm{so}/2)\sum_i \bm L_i
\cdot \bm \sigma_i$ as a perturbation for each band obtained above.
The spin-orbit interaction energy $\lambda_\mathrm{so}$ is assumed
smaller than either the ISB energy $\gamma$ or the bandwidth
$\Delta$. Employing the standard degenerate perturbation theory for
the two spin states at a given $\v k$-vector,  we find the resulting
matrix elements $\langle n, \v k, s | H_\mathrm{so} |n, \v k, s'
\rangle$ ($s, s' = \uparrow, \downarrow$) form a $2\times2$
Rashba-type Hamiltonian matrix

\ba \pm {3\over 2}\lambda_\mathrm{so} {\gamma  \over \Delta }
\hat{z}\cdot \v k \times \bm \sigma \label{eq:Rashba-H}\ea
for bands $n=2$ and $n=3$, respectively. There is no Rashba
splitting for band 1 within the degenerate perturbation theory.
Combined with the analysis of chiral OAM, we obtain the following
hierarchy of chiral angular momenta: First, there is chiral OAM
arising from ISB alone. At the next level, when a small SOI is
present, each spin-degenerate OAM-carrying band is split into a pair
of bands carrying opposite chiral SAMs. Both chiral structures are
embodied in Dirac-like effective Hamiltonians, with the energy scale
for chiral OAM dominant over that of chiral SAM by the factor
$\Delta/\lambda_\mathrm{so}$. Chiral OAM, unlike SAM, can be
substantial for surfaces consisting of light elements, and its
strength controlled by applying an inversion-symmetry-breaking
electric field externally.

\begin{figure}[ht]
\includegraphics[width=85mm]{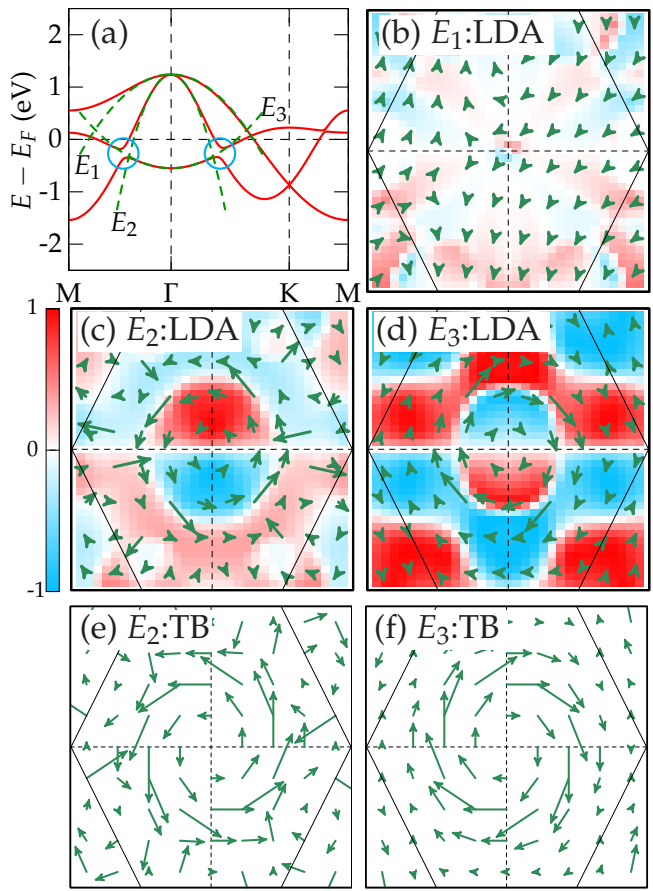}
\caption{OAM and CD from first-principles and tight-binding
calculations of Bi monolayer without SOI. (a) LDA band structure for
Bi monolayer of triangular lattice structure without SOI.
Perpendicular electric field of 3V/\AA~was imposed externally. Three
dashed curves represent the tight-binding energy dispersions around
the $\Gamma$ point. (b)-(d) OAM vectors (green arrows) and CD signals
(color backgrounds) for the three bands from the highest ($E_1$) to
lowest ($E_3$) energies over the whole Brillouin zone marked by a
solid hexagon. Largest OAM has magnitude $\approx 1\hbar$ for bands
$E_2$ and $E_3$. Incoming photon direction $\v k_\mathrm{ph} =
+\hat{x}$ is assumed for the CD calculation. (e)-(f) OAM obtained
from the tight-binding calculation for two main OAM-carrying bands
$E_2$ and $E_3$. } \label{fig:LDA-CD-woSO}
\end{figure}

To ensure that chiral OAM exists in a more realistic calculation, we
performed first-principles local-density approximation (LDA)
calculation for a Bi single layer forming a triangular lattice. The
choice is inspired by Bi being a proto-typical $p$-orbital band
material. An external electric field of 3V/\AA~ perpendicular to the
layer was imposed by hand to mimic the surface potential gradient
without having the complication of dealing with the bulk states. The
physically more relevant case of Bi layer is also
considered\cite{liu}, with LDA results easily adaptable to the
single-layer case discussed below. To emphasize the relevance of ISB
we chose to investigate the spin-degenerate by turning off SOI in
the LDA calculation. The resulting electronic structure for spinless
case consisting of three $p$-orbital-derived bands is shown in
Fig.~\ref{fig:LDA-CD-woSO}(a). As the external electric field is
turned on, a level repulsion between the middle ($E_2$ in Fig. 1(a))
and the bottom ($E_3$ in Fig. 1(a)) band occurs as indicated by
circles in Fig.~\ref{fig:LDA-CD-woSO}(a). These two bands exhibit
the chiral OAM patterns with the maximum OAM vector $|\langle \bm L
\rangle | \approx 0.96\hbar$ as shown in
Fig.~\ref{fig:LDA-CD-woSO}(c) and 1(d), while the third one, shown
in Fig. 1(b), carries much less OAM around the $\Gamma$ point. The
OAM chiralities of the two bands are opposite, in accordance with
the previous TB analysis. An excellent fit of the LDA band structure
near the $\Gamma$ point was possible with the TB parameters
$V_1=-0.725$ eV, $V_2=-0.11$ eV and $\gamma=0.2623$ eV (Fig. 1(a)).
The good fit by the TB model is a testament to the negligible
crystal field effect in the elemental Bi. OAM values obtained by the
TB analysis are also in excellent quantitative accord with the LDA
results as can be seen by comparing Fig.
\ref{fig:LDA-CD-woSO}(c)-(d) (LDA) to Fig.
\ref{fig:LDA-CD-woSO}(e)-(f) (TB). The OAM magnitude is seen to
decrease continuously upon approaching the $\Gamma$ point in the LDA
calculation (Fig. 1(c) and 1(d)) as predicted by the TB calculation,
Eq. (\ref{eq:perturbative-OAM}).

Having established theoretically the existence of chiral OAM in
inversion-asymmetric bands by a number of methods, we turn to the
question of its detection. Spin- and angle-resolved photoemission
spectroscopy (SARPES) has served to identify the chiral spin
structure of the surface bands in the past\cite{SARPES,kimura}. A
similar chiral structure for OAM as demonstrated here cannot,
however, be detected by the same probe since chiral OAM exists even
when SOI is very weak and spin degeneracy is nearly perfect. Circular
dichroism (CD) refers to phenomena in which the physical response of
a system to probing light depends systematically on the light
polarization being left-circularly-polarized (LCP) or
right-circularly-polarized (RCP). In ARPES experiment, in particular,
incident lights of opposite helicities (RCP vs. LCP) might give rise
to different scattering intensities of photo-electrons. It can be
shown, in fact, that such difference is a consequence of finite OAM
polarization in $\v k$-space and can be used to detect its existence.

One can formally define the CD-ARPES signal $D(\v k)$ as

\ba D(\v k) &=& {\sum_\sigma  \bigl(  I^{\rm{RCP}}_\sigma (\v k) -
I^{\rm{LCP}}_\sigma (\v k) \bigr) \over \sum_\sigma \bigl(
I^{\rm{RCP}}_\sigma (\v k) + I^{\rm{LCP}}_\sigma (\v k)\bigr)} .
\label{eq:CD-formula}\ea
Here $I_\sigma (\v k)$ refers to the probability of scattering from
an initial occupied state at momentum $\v k$ to a final
photo-electron state of spin orientation $\sigma$. Since we have
spin-integrated ARPES in mind, the final-state spin is summed.
Coupling to electromagnetic fields gives rise to a perturbed
Hamiltonian $ H_1 \sim \v p \cdot \v A$ ($\v p$=momentum operator,
$\v A$=vector potential). The transition amplitude from the initial
state $|I\rangle$ to a final state $|F\rangle$ is proportional to
$\langle F | H_1 | I \rangle \sim \langle F|\v r \cdot \v A |I\rangle
$, according to standard theory. Taking a plane-wave form for the
final state wave function $\psi_F (\v r) = e^{i \v k_F \cdot \v r}$,
where $\v k_F$ is the wave vector of the final-state photo-electron,
and using the initial state wave function obtained from LDA, we can
calculate the transition probability for RCP ($\v A$) and LCP ($\v
A^*$) incoming lights to get $D(\v k)$. The in-plane component of $\v
k_F$ matches the momentum of band electrons (higher-order Umklapp
processes are neglected), while the normal component $k_F^z$ can be
deduced from energy conservation. 

The incident photon direction $\hat{k}_\mathrm{ph}$ is chosen with
its in-plane component along $+\hat{x}$ axis in calculating $D(\v k)$
for Bi monolayer band structure.  The results are color-coded and
superimposed on the OAM patterns in Fig. 1(b)-(d). In the case of Bi
monolayer the CD intensity proved largely insensitive to the choice
of $k_F^z$ value, so we used $k_F^z = 0$ for the displayed result.
One can make sense of the CD color patterns from a simple angular
momentum conservation argument when $\v k$ is close to the $\Gamma$
point. Defining the projection of the electron angular momentum
operator $\bm L$ onto the photon direction as
$\hat{k}_\mathrm{ph}\cdot \bm L$, incident RCP light would increase
it by $+\hbar$ whereas LCP light reduces it by $\hbar$. Therefore,
loosely speaking, initial states with $\langle \bm L \rangle \propto
-\hat{x}$ (\textit{l}=-1) is coupled to $l=0$ ($l=-2$) channel of the
final state by the RCP (LCP) light. Expansion of the final plane-wave
state in terms of spherical harmonics produces a larger weight for
$l=0$ than $l=-2$, thus $D(\v k)>0$ for such an initial state.
Following such consideration leads to the result $D(\v k) \propto
\hat{k}_\mathrm{ph}\cdot \langle \v k | \bm L | \v k \rangle$.

In Fig. 1(c)-(d), one finds that such elementary consideration to be
indeed in accord with the CD patterns near the $\Gamma$ point. Far
from it, though, the color pattern is no longer consistent with it
and even gives out the opposite sign, implying that effects besides
the conservation rule play a crucial role in determining the overall
sign of $D(\v k)$. In principle, however, all such effects are
captured in the LDA calculation and should be compatible with ARPES
experiment. We emphasize that CD signal vanishes where OAM
disappears, leaving little doubt that the measurement of CD is a
direct indicator of local OAM in momentum space. On the other hand,
the existence of spin polarization in the band will not be captured
by the proposed method even when it is prominent.

\begin{figure}[ht]
\includegraphics[width=85mm]{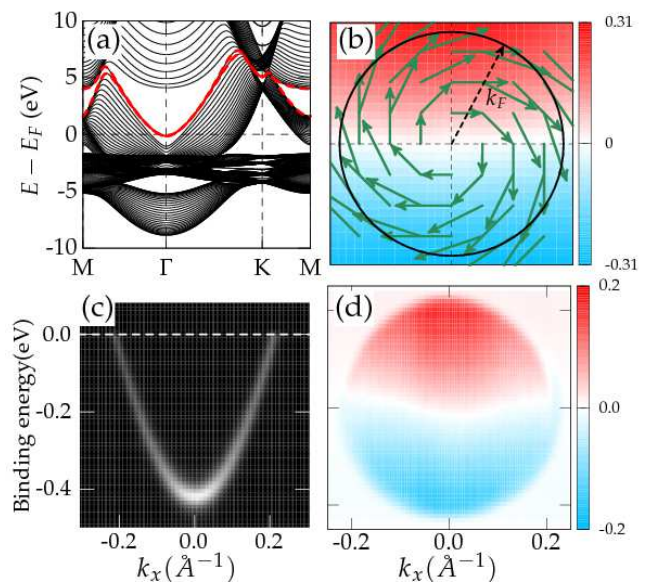}
\caption{OAM and CD of Cu surface band. (a) First-principles
electronic band structure of Cu(111) 30-layer slab. Red dashed lines
indicate the two surface bands. (b) Calculated OAM and CD (assuming
the in-plane incident photon direction $ \hat{\v k}_\mathrm{ph}
\propto +\hat{x}$) corresponding to outer energy surface bands.
Inner surface band shows basically the same OAM and CD patterns.
Maximum OAM vector is $\sim 0.07\hbar$. (c) ARPES-measured surface
band of Cu along $k_y=0$. (d) Measured CD-ARPES result $D(\v k)$.
Incident photon direction has in-plane component along $+\hat{x}$.
The momentum window in Fig. (b) matches that of Fig. (d). }
\label{fig:Cu}
\end{figure}

Finally, we performed CD-ARPES experiment on Cu surface with a view
to confirm the existence of chiral OAM on a realistic material with
a small SOI. Because SOI is very small in Cu, it can be safely said
that chiral OAM structure is, without question, due to the ISB
mechanism proposed here. The calculated band structure of 30 Cu(111)
layers is shown in Fig.~\ref{fig:Cu}(a). Two surface-derived bands
are identified as red curves. In the case of Cu calculation the
inversion symmetry-breaking electric field is generated
spontaneously through self-consistent electronic structure
calculation, and not imposed externally as in the Bi monolayer case.
Despite these differences we found a clear, chiral OAM pattern
around the $\Gamma$ point as shown in Fig. ~\ref{fig:Cu}(b). Even
though the make-up of the surface bands receive substantial
contributions from both $p$- and $d$-orbitals, OAM arises
predominantly from the $d$-orbital components. Using Eq.
(\ref{eq:CD-formula}) we can also calculate $D(\v k)$ for Cu.
Contrary to the model Bi calculation, $D(\v k)$ varies significantly
with the choice of the final-state momentum $k_F^z$. For Fig.
\ref{fig:Cu}(b) we used $k_F^z = 2.27$\AA$^{-1}$ consistent with the
photo-electron energy in the actual experiment described in the next
paragraph. The sign of $D(\v k)$ is opposite to the anticipation
from the angular momentum conservation rule near the $\Gamma$ point.
It agrees, however, with the following dichroism experiment
performed on Cu surface.

The single, sharp experimental Cu(111) surface band shown in Fig.
\ref{fig:Cu}(c) actually consists of two surface bands with opposite
spins as found in LDA calculation, unresolved because of the small
energy spacing between them. We take the ARPES data twice, one with
RCP and the second with LCP lights, and obtain $D(\bf k)$ according
to Eq. (\ref{eq:CD-formula}). The results, plotted in Fig.
\ref{fig:Cu}(d), cover the small occupied region in Brillouin zone
with the Fermi momentum $\approx 0.2$\AA$^{-1}$. The CD-ARPES
experiment shows pronounced CD behavior that reaches a maximum value
of $\approx 20\%$ at the Fermi surface, where the OAM is also
expected to be the largest. Most importantly, the sign of CD matches
that in the LDA calculation, which is a highly non-trivial result
given the reversal of the CD sign discussed earlier. The calculated
$D(\v k)$ varies between -0.3 and +0.3, somewhat greater than the
20\% variation found experimentally. Thermal effects in the
experiment may have contributed to the smearing of CD signal.

Our prediction, and its observation on Cu surface, of chiral OAM is a
direct analogue of the well-known chiral SAM from Rashba-split
surface bands. Likewise, much of the physical consequences of chiral
SAM is expected to have orbital analogues. For instance, spin
polarization can drive electrical current through spin-momentum
locking in a phenomenon known as the spin Galvanic
effect\cite{ganichev}. Its inverse, the spontaneous spin polarization
in the current-carrying state of Rashba-split bands, was proposed
theoretically by Edelstein\cite{edelstein} and confirmed
experimentally in recent years\cite{chernyshov}. Having identified
OAM as the more fundamental physical quantity associated with ISB, we
propose that both these spin-related effects must have orbital
analogues.

\acknowledgments This work is supported by Mid-career Researcher
Program No. 2011-0015631 (JHH), the KICOS through Grant No.
K20602000008 (CK), and R17-2008-033-01000-0 (JY). We acknowledge
fruitful conversations with Seung Ryong Park.

\end{document}